\documentclass[aps,prb,twocolumn,showpacs,superscriptaddress]{revtex4-1}\begin{tiny}\end{tiny}
\usepackage{longtable}
\usepackage{amsfonts}
\usepackage{graphicx}
\usepackage{setspace}
\begin{document}
\title{Low temperature semiconductor band gap thermal shifts: \\
$T^4$ shifts from ordinary acoustic and $T^2$ from piezo-acoustic coupling}
\author{Philip B. Allen and Jean Paul Nery}
\email{philip.allen@stonybrook.edu}
\email{jeanpaul240@gmail.com}
\affiliation{Physics and Astronomy Department, Stony Brook University, Stony Brook, NY 11794-3800, USA}
\date{\today}
\begin{abstract}
At low temperature, the experimental gap of silicon decreases as $E_g(T)=E_g(0)-AT^4$.   
The main reason is electron-phonon renormalization. The physics behind the $T^4$-power law
is more complex than has been realized.  Renormalization at low $T$ by intraband scattering requires 
a non-adiabatic treatment, in order to correctly include acoustic phonons, and avoid divergences from
piezo-acoustic phonon interactions.  The result is an unexpected low $T$ term $E_g(0)+A^\prime T^p$
with positive coefficient $A^\prime$, and power $p=4$ for non-piezoelectric materials, and
power $p=2$ for piezoelectric materials.  The acoustic phonons in piezoelectric 
semiconductors generate a piezoelectric field, modifying the electron-phonon coupling.  
However, at higher $T$, thermally excited acoustic phonons of energy  $\hbar v_s q$ 
and intraband excitation energies $\epsilon_q-\epsilon_0 = \hbar^2 q^2 /2m^\ast$
become comparable in size.  Above this temperature, the low $q$ and higher $q$
intraband acoustic phonon contributions to $T^p$ rapidly cancel, leaving little thermal
effect.  Then the contribution from interband scattering by acoustic phonons is dominant.
This has the power law $T^4$ for both non-piezoelectric and piezoelectric semiconductors.
The shift can then have either sign, but usually reduces the size of gaps as $T$ increases.
It arises after cancelation of the $T^2$ terms that appear separately in Debye-Waller 
and Fan parts of the acoustic phonon interband renormalization.  The cancellation occurs because
of the acoustic sum rule.  
\end{abstract}

%
%
\maketitle

\section{Introduction} \label{sec:intro}
 
Electron bands in crystals have temperature-dependent energies\cite{Cardona2005}.  Typical values
at room temperature and above are $E_g(T)-E_g(0)\sim(2-5)k_B T$.
There are separate
contributions from thermal expansion and electron-phonon interactions.  For heavy elements, these
contributions\cite{Monserrat2016} are similar in size, but for lighter mass elements, 
electron-phonon effects are significantly bigger than thermal expansion effects.
The subject has been studied, by second-order perturbation theory, for a long time.
Recent progress in computational theory \cite{Giustino_Louie_Cohen, X2, X3, X4, Ponce_Gonze, Calandra, Verdi,
Zacharias, Gibbs, Ponce_Gonze2, 
Antonius_Gonze, metal1, metal2, Brown} has enabled microscopic calculations.  

These {\it ab initio} calculations have difficulties with energy denominators 
$\epsilon_{\mathbf{k},n}-\epsilon_{\mathbf{k}-\mathbf{q}, n^\prime} \pm\hbar\omega_{\mathbf{q}j}$,
when they are small.  Fortunately it is often possible to drop the phonon energy $\pm\hbar\omega_{\mathbf{q}j}$
compared to the electron energy difference, replacing it with an {\it ad hoc} \
$i\Delta$ with $\Delta\sim$0.1 eV, designed to smooth
out an integrable singularity.  This is an adiabatic approximation, where electrons
are unaware of the time-dependence of vibrational motion.  Unfortunately, this approximation
is sometimes impossible.  Intraband scattering by polar optical modes (the Fr\"ohlich problem)
is the known example\cite{Ponce_Gonze}.  A divergence (a non-integrable singularity) 
occurs in adiabatic approximation.  Adding $i\Delta$ gives an incorrect answer\cite{JPNPBA} that
depends on $|\Delta|$.  The true Fr\"ohlich answer is found by a principle-parts
integration over the integrable singularity that occurs in the correct non-adiabatic treatment.  Here we find
that at very low $T$, intraband scattering by acoustic phonons also requires a non-adiabatic treatment.

In a landmark paper, Cardona {\it et al.} \cite{CMT} measured the band gap of silicon to extraordinary accuracy at low $T$.
Their result, for $1{\rm K}<T<4{\rm K}$, is  $E_g\approx1.15{\rm eV}-250{\rm eV}(T/\Theta_{\rm D})^4$, where the 
Debye temperature is $\Theta_{\rm D}$=645K. The 4$^{\rm{th}}$ power of $T$ behavior disagreed 
with earlier fits on other materials
\cite{Passler} which gave powers in the range $2-3.3$.  Cardona {\it et al.} provided a correct qualitative 
argument favoring $T^4$.  It has been
repeated in more detail since then \cite{Monserrat}.  Here we show that although $T^4$ is supported by theory
when $T$ is not too low, there is actually a lower $T$ regime where the temperature shift always increases
the gap, with a power law $T^4$ in  non-piezoelectric crystals like diamond and rocksalt,
and $T^2$ in piezoelectric semiconductors like zincblende.  The very low $T$ behavior comes from non-adiabatic
effects involving intraband virtual emission and absorption of acoustic modes.  
If the material is piezoelectric, then an (incorrect) adiabatic treatment of the piezo-acoustic intraband scattering
diverges at all $T$.  However, except at very low $T$, it is accurate to drop the (correct non-adiabatic) piezo-electric
part of the acoustic phonon coupling.  At very low
$T$, this term dominates, giving a $T^2$ shift.

The higher $T$ thermal shifts,
where the gap decreases like $-A(T/\Theta_{\rm D})^4$, come from interband virtual scattering.  We show that
the $T^4$ behavior results from a cancellation to order $q^2$ between the interband
Fan-type \cite{Fan} terms and the  Debye-Waller-type \cite{Antoncik} terms.  The cancellation holds in the regime
where the adiabatic treatment is accurate to order $q^2$.  The cancellation follows 
from the acoustic sum rule \cite{AH}. 

In section \ref{sec:ge}, the underlying theory is reviewed.  The new results for small-$\mathbf{q}$
acoustic coupling are explained in Sec. \ref{sec:NA}.  The higher temperature $T^4$ result, from
interband virtual scattering with cancellation of separate $T^2$ contributions, is derived in Sec. \ref{sec:IB}.
Debye-model estimates, and ideas for improving numerical codes, are also given.  Finally,
Sec. \ref{sec:C} contains a summary.

\section{general equations} \label{sec:ge}

It is worth beginning with the influence of thermal expansion on band energies.  This is partly because the equation has close
similarities to the electron-phonon results to be discussed, but also because it illustrates nicely the simplicity that
turns out to be elusive in the case of acoustic phonon contributions to electron energy renormalization.  
Standard quasiharmonic theory \cite{Gruneisen,Ashcroft} gives for the shift of volume,
\begin{equation}
\frac{\Delta V}{V_0} = \frac{1}{NB_0 V_0}\sum_{\mathbf{q}s} \hbar\omega_{\mathbf{q}s} \gamma_{\mathbf{q}s}
(n_{\mathbf{q}s} + 1/2).
\label{eq:thexp}
\end{equation}
Here $\gamma_{\mathbf{q}s}$ is the mode Gr\"uneisen parameter $-(V_0/\omega_{\mathbf{q}s})
(\partial\omega_{\mathbf{q}s}/\partial V)_0$,
and $\omega_{\mathbf{q}s}$ is the frequency of a phonon mode, with thermal occupancy given by the Bose-Einstein
distribution $n_{\mathbf{q}s}$. 
$B_0$ is the bulk modulus, $V_0$ is the volume of the unit cell, $NV_0$ is the volume of the sample, and subscripts $0$
denote values computed for the frozen-lattice (Born-Oppenheimer) ground state.
Now let $D_{\alpha\beta}(\mathbf{k}n)$ denote the deformation potential \cite{Bardeen,Cardona}
$\partial \epsilon_{\mathbf{k}n}/\partial \epsilon_{\alpha\beta}$,
the rate of shift of an electron Bloch energy $\epsilon_{\mathbf{k}n}$ per unit strain $\epsilon_{\alpha\beta}$.
In a cubic material, a symmetric electron state at $\mathbf{k}=0$ has 
$V_0(\partial\epsilon_{\mathbf{k}=0,n}/\partial V)_0 = D_{\alpha\alpha}(0n)\equiv D$,
valid for any direction $\alpha$.  The electron energy is then $E_{\mathbf{k}n} = \epsilon_{\mathbf{k}n}+\delta_{\mathbf{k}n}$,
and $\delta_{\mathbf{0}n}=(D/B_0 V_0)\sum \hbar\omega_{\mathbf{k}^\prime s}\gamma_{\mathbf{k}^\prime s}
(n_{\mathbf{k}^\prime s}+1/2)$.
Temporarily keeping only the thermal part $\delta_{\mathbf{k}n}(T)-\delta_{\mathbf{k}n}(0)$, that is, dropping the $1/2$, 
the Debye model gives
\begin{eqnarray}
[\delta_{\mathbf{k}n}(T)-\delta_{\mathbf{k}n}(0)]_{\mathbf{k}=0} &=& \frac{9D}{B_0 V_0} \bar{\gamma}\hbar\omega_D \left(\frac{T}{\Theta_{\rm D}}\right)^4 \nonumber \\
&\times&\int_0^{\Theta_{\rm D}/T} dx\frac{x^3}{e^x-1},
\label{eq:eshthexp}
\end{eqnarray}
where the Gr\"uneisen parameter is assumed constant, $\gamma_{\mathbf{q}s}\rightarrow\bar{\gamma}$.  
At low $T$, the upper limit of the integral 
is $\infty$ and the result is $\delta(T)-\delta(0) = (3\pi^4 /5)(D/B_0 V_0) \bar{\gamma}\hbar\omega_D (T/\Theta_{\rm D})^4$.
It is natural to expect that acoustic phonon contributions to the electron-phonon part of the
energy shift should have simple power laws given by closely related formulas.  To estimate the size
of the thermal expansion term, $|D/B_0 V_0|$ is of order 1,  and $\bar{\gamma}\hbar\omega_D$ is commonly about
0.1eV, so $|\delta(T)-\delta(0)| \sim6(T/\Theta_{\rm D})^4$eV.  To be more specific for silicon, the deformation potential 
for the gap\cite{Bardeen} ($E_c -E_v$) is
$\sim - 30$eV.  Then the prefactor of $(T/\Theta_{\rm D})^4$ is $\sim-$8eV (taking $\bar{\gamma}\sim 1$),
which is small compared to the measured \cite{CMT} prefactor, $\sim-$250eV.  At higher $T$ ($>\Theta_{\rm D}$)
where Eq.(\ref{eq:eshthexp}) is linear in $T$, the thermal expansion contribution is typically $\sim$25\% of
the total thermal shift.  Optic phonons now contribute equally strongly as acoustic, both to the
thermal expansion effect and to the electron-phonon renormalization, which is also linear in $T$.  
It is of course common for Eq.(\ref{eq:eshthexp}) to misrepresent
the full $T$ dependence from expansion.  Some crystals ({\it e.g.} silicon) have
sign changes of $dV/dT$ as $T$ increases.  These occur because $\gamma_{\mathbf{q}s}$ can vary in sign for different 
regions of the phonon spectrum.  But the $T^4$ power law for $\Delta V$ is secure at low $T$, whereas the 
electron-phonon contributions to $\Delta E$ have more fundamental issues, to be explained in Sec. \ref{sec:NA}. 

Formulas for the electron-phonon renormalization are available in literature \cite{X3,Allen}.  It is helpful
to split the answer into the intraband term and the rest (interband).  The intraband term
needs to be treated without making the adiabatic approximation, but the interband term is
accurately treated by neglecting the phonon frequency in the denominator.
\begin{equation}
\Delta E_{\mathbf{k}n}^{\rm ep}\equiv (E_{\mathbf{k}n} - \epsilon_{\mathbf{k}n})_{\rm ep} =
\Delta_{\mathbf{k}n}^{\rm non-adia}+\Delta_{\mathbf{k}n}^{\rm inter}.
\label{eq:defad}
\end{equation}
The non-adiabatic formula, for the shift in energy of an electron state $\mathbf{k}n$, is
\begin{eqnarray} 
&&\Delta_{\mathbf{k}n}^{\rm non-adia}=\frac{{\cal R}e}{N}\sum_{\mathbf{q}s}
 | \langle\mathbf{k}+\mathbf{q}n| V_1
(\mathbf{q}s)|\mathbf{k}n\rangle|^2 \times \nonumber  \\
&&\left[ \frac{1+n_{-\mathbf{q}s}-f_{\mathbf{k}+\mathbf{q}n}} 
{\epsilon_{\mathbf{k}n}-\epsilon_{\mathbf{k}+\mathbf{q}n}-\hbar\omega_{-\mathbf{q}s}+i\eta}
+\frac{n_{\mathbf{q}s}+f_{\mathbf{k}+\mathbf{q}n}} 
{\epsilon_{\mathbf{k}n}-\epsilon_{\mathbf{k}+\mathbf{q}n}+\hbar\omega_{\mathbf{q}s}+i\eta}  \right]. \nonumber \\
\label{eq:DeltaNon}
\end{eqnarray}
Here $f_{\mathbf{k}+\mathbf{q}n}$ is the Fermi-Dirac occupation factor for the intermediate
electron state $\mathbf{k}+\mathbf{q}n$, 
$n_{\mathbf{q}s}$ is the Bose-Einstein thermal occupation
of the phonon state $\mathbf{q}s$ of energy $\hbar\omega_{\mathbf{q}s}$, and $i\eta$ is an infinitesimal
imaginary shift.  The operator $V_1(\mathbf{q}s)$ is the first-order electron-phonon
interaction, $(\partial V/\partial u_{\mathbf{q}s})u_{\mathbf{q}s}$.
Taking the real part means a principal-part treatment of the zero in the denominator.
This formula, first given by Fan \cite{Fan}, is most easily derived from a diagrammatic treatment of the electron self-energy
as formulated by Migdal \cite{Migdal} and Eliashberg \cite{Eliashberg}.  We have assumed here that the state 
$|\mathbf{k}n\rangle$ under consideration is non-degenerate.  The degenerate case has been discussed
by Trebin and R\"ossler\cite{Trebin}.


The rest is adiabatic, and the formula is
\begin{eqnarray} 
&&\Delta_{\mathbf{k}n}^{\rm inter}=\sum_{\ell i \alpha, m j \beta} \left\{ \sum_{\mathbf{q}} \sum_{n^\prime}^{\ne n}
\right.  \nonumber \\
 &&\frac{ \langle\mathbf{k}| \frac{\partial V}{\partial u_{\ell i \alpha}}|\mathbf{k}+\mathbf{q}n^\prime\rangle 
  \langle\mathbf{k}+\mathbf{q}n^\prime| \frac{\partial V}{\partial u_{m j \beta}}|\mathbf{k}n\rangle} 
{\epsilon_{\mathbf{k}n}-\epsilon_{\mathbf{k}+\mathbf{q}n^\prime}}
 \nonumber \\
&&\left.  \ \ \ \ \ \ \ \ +  \frac{1}{2}
\langle \mathbf{k}n |  \frac{\partial^2 V}{\partial u_{\ell i \alpha}\partial u_{mj\beta}} |\mathbf{k}n\rangle \right\}
 \langle u_{\ell i\alpha}u_{mj\beta}\rangle.
\label{eq:DeltaAd}
\end{eqnarray}
This is written in terms of the real space lattice displacements $u_{\ell i \alpha}$, for
reasons related to the ``acoustic sum rule'' which will appear soon. The indices of summation
$\ell$ and $m$ enumerate the unit cells located at $\mathbf{R}_\ell$ and $\mathbf{R}_m$;
indices $i$ and $j$ go over the atoms within the unit cell; $\alpha$ and $\beta$ are Cartesian directions.
The meaning of Eq.(\ref{eq:DeltaAd}) is that since ions move slowly compared to electrons, their displacements
can be approximated as static.  For any particular static displacement, the energy shift is computed to second
order by standard perturbation theory.  Finally, the result is averaged over the thermal distribution of displacements
using harmonic lattice dynamics.

The lattice displacement $u_{\ell i\alpha}$ is
\begin{equation}
u_{\ell i\alpha} = \sum_{\mathbf{q}s}(\hbar/2M_i N\omega_{\mathbf{q}s})^{1/2}
 \epsilon_{i\alpha}(\mathbf{q}s) \exp(i\mathbf{q}\cdot\mathbf{R}_\ell)
\phi_{\mathbf{q}s}
\label{eq:udisp},
\end{equation}
where $\epsilon_{i\alpha}(\mathbf{q}s)$ is the polarization vector of mode $\mathbf{q}s$,  normalized 
by $\sum |\epsilon_{i\alpha}(\mathbf{q}s)|^2=1$, when summed over $i\alpha$.  The operator 
$\phi_{\mathbf{q}s}$ equals $a_{\mathbf{q}s}+a^\dagger_{-\mathbf{q}s}$, where $a$ and $a^\dagger$ are phonon
destruction and creation operators.  Using this, it is easy to convert Eq.(\ref{eq:DeltaAd}) to the reciprocal space
version similar to Eq.(\ref{eq:DeltaNon}).

The first term in Eq.(\ref{eq:DeltaAd}) is the interband generalization (omitted in Fan's original paper \cite{Fan}) of the Fan term. 
and the second (first given by Antoncik \cite{Antoncik}) is the ``Debye-Waller'' term.  
The Fan-type interband terms have the same structure as the non-adiabatic part, Eq.(\ref{eq:DeltaNon}), except
that the intermediate electron state $\mathbf{k}+\mathbf{q}n^\prime$ is in a different band $n^\prime\ne n$, 
and the phonon frequencies $\pm\hbar\omega_{\mathbf{q}s}$ are omitted (for convenience)
from the denominators.  The reason why this omission is safe is that the interband energy differences
$|\epsilon_{\mathbf{k}n}-\epsilon_{\mathbf{k}+\mathbf{q}n^\prime}|$ are typically at least 10 (and usually more) times
bigger than  $\hbar\omega_{\mathbf{q}s}$.  When dealing with electron states $\mathbf{k}n$ not at band extrema,
there will always be surfaces in $\mathbf{k}$-space with zero denominators.  Omitting $\pm\hbar\omega_{\mathbf{q}s}$
merely shifts the position of these surfaces.  The principal parts evaluation of integrals, over $\mathbf{k}$-space volumes
that contain such singularity surfaces, gives finite answers that are usually not large, 
and not expected to change much when $\pm\hbar\omega_{\mathbf{q}s}$
is included.  These arguments fail near band extrema for the intraband $n=n^\prime$ terms.  This is why
a non-adiabatic treatment is needed for the intraband case.  They also fail in metals at low $T$ for a different reason,
namely that we care most about states lying exactly in the region of the Fermi energy.  Then the singularity surface is close
to the Fermi surface, where the intraband Fermi-Dirac function $f_{\mathbf{k}+\mathbf{q}n}$ has sharp variation
on the small energy scale $k_B T$.  This destroys the simple smallness of the principal parts integration.

Allen and Heine \cite{AH} gave a sum rule which clarified the need for both Fan and Debye-Waller contributions,
and showed how they are linked,
\begin{eqnarray}
0&=&\sum_{\ell i\alpha,m j \beta} \left[ \sum_{ n^\prime}^{\ne n} 
\frac{ \langle\mathbf{k}n | \frac{\partial V}{\partial u_{\ell i\alpha}}|\mathbf{k} n^\prime\rangle
\langle\mathbf{k} n^\prime| \frac{\partial V}{\partial u_{mj\beta}}|\mathbf{k}n\rangle} 
{\epsilon_{\mathbf{k}n}-\epsilon_{\mathbf{k} n^\prime}} \right. \nonumber \\
 && \ \ \ \ \ +\frac{1}{2} \left. \langle\mathbf{k}n| \frac{\partial^2 V}{\partial u_{\ell i\alpha}\partial u_{mj\beta}} |\mathbf{k}n\rangle
\right] A_\alpha A_\beta.
\label{eq:sumrule}
\end{eqnarray}
This equation simply says that, when every atom is displaced statically (hence $\omega_{\mathbf{q}s}$ is set to 
zero in the denominator),  by the same arbitrary constant vector $\mathbf{A}$, 
there is no shift of any electron eigen-energy.  The displacements $A_\alpha$, $A_\beta$
must be independent of $\ell m$ and $ij$.  

Numerical studies using phonons and coupling from density functional theory (DFT) have become very 
powerful.  A few comments are appropriate.  It has been found useful to add $i\Delta$ to the energy denominators
in Eq.(\ref{eq:DeltaAd}), where $\Delta\sim$0.1eV is a typical choice.  This is a sensible way to avoid the difficulties
of principal parts integration when there is a singularity surface.  The {\it ad hoc} cure works well, but cannot give 
correct low $T$ power laws that
arise from acoustic phonons.  This is a small effect which is totally unimportant at higher $T$.  Another deficiency
of the $i\Delta$ cure is to distort the polaronic contribution from small $q$ polar optical modes which have
a particularly strong coupling to electrons.  In another paper \cite{JPNPBA} we derive an approximate correction
to deal with this. 
Eqs.(\ref{eq:defad}-\ref{eq:sumrule}) contain the ingredients needed for our analysis.  

\section{Nonadiabatic effects of acoustic phonons} \label{sec:NA}

To simplify things, we are most interested in band gaps. For both valence band maxima
and conduction band minima, the energy differences $\epsilon_{\mathbf{k}+\mathbf{q}}-\epsilon_{\mathbf{k}}$
have the form $\hbar^2 q^2 /2m^\ast$ (in effective mass approximation).
Often these are at wavevector $\mathbf{k}=0$, and when they are not, we will
simplify the notation by denoting the site of the band extremum as $\mathbf{k}=0$.  Then
the ``non-adiabatic acoustic" (N,A) contribution $\Delta\epsilon$ to the shift of a state at a band edge can be written
\begin{eqnarray}
&&\Delta_{\mathbf{k}=0}^{\rm N,A} =\sum_{\mathbf{q}s} | \langle\mathbf{k+q}=
\mathbf{q}| V_1(\mathbf{q}s)|\mathbf{k}=0\rangle|^2 \nonumber \\
&& \ \ \times\left[\frac{1+n_{-\mathbf{q}s}} {-\hbar^2 q^2 /2m^\ast -\hbar v_s q}
+ \frac{n_{\mathbf{q}s}} {-\hbar^2 q^2 /2m^\ast +\hbar v_s q} \right].
\label{eq:deltac}
\end{eqnarray}
The band index $n$ has been dropped.  As written,
the formula applies to an electron state at the bottom of the conduction band.
For a hole state at the top of the valence band, it is necessary to replace $1+n_{-\mathbf{q}s}$ by $n_{\mathbf{q}s}$,
and vice versa.  The sign also has to be changed, if we want the energy shift of the electron state
at the top of the valence band, rather than the shift of the hole energy.  
No excited or doped electrons in the conduction band, or holes in the valence
band are present, so the Fermi factors have also been dropped.  The sum over $\mathbf{q}$ must be restricted to small
wavevectors (typically 10\% of the distance to the Brillouin zone boundary) where the effective mass approximation
for the band energy can be trusted.

\subsection{Piezo-acoustic coupling} \label{sec:PA}

Piezoelectric materials acquire a polarization $\mathbf{P}$ proportional to strain.  The linear relation is
$P_\alpha = e_{\alpha\beta\gamma}\epsilon_{\beta\gamma}$, where the third rank tensor $e_{\alpha\beta\gamma}$ is
the piezoelectric tensor, and the second rank tensor $\epsilon_{\beta\gamma}$ gives the strain.
The piezoelectric tensor can be computed \cite{Vanderbilt}.
Zincblende structure is the simplest piezo-electric semiconductor structure, with only a single 
piezo-electric constant, $e_{xyz}=e_{yzx}=e_{zxy}=-e_{xzy}$ {\it etc.}
When numbers are needed, we use the metastable zincblende version of GaN (denoted c-GaN) as the example.
In zincblende crystals, a shear strain in the $xy$-plane creates a polarization, and
an $\mathbf{E}-$field, in the $z$ direction.

The coupled system of an electron and piezo-active acoustic phonons is known as the piezo-polaron \cite{Hutson}.  
The small $q$ intraband piezoelectric matrix element is \cite{Cardona}
\begin{equation}
\langle\mathbf{q}| V_1(\mathbf{q}s)|\mathbf{0}\rangle=
g_{\mathrm{piezo}}=- \frac{e}{4 \pi \tilde{\varepsilon}_0} \frac{q_\alpha e_{\alpha\beta\gamma}
(iq_\beta u_\gamma)}{q^2 \varepsilon_{\infty}}
\label{eq:pcoup}
\end{equation}
where $\tilde{\varepsilon}_0$ is the vacuum permittivity, and
$u_\gamma$ is the acoustic phonon amplitude.  In zincblende, the angular average of the squared matrix element
is 
\begin{eqnarray}
\langle | g_{\mathrm{piezo}}|^2\rangle&=&
\left( \frac{e^2}{4 \pi \tilde{\varepsilon}_0 \epsilon_\infty a}\right)^2 \left( \frac{e_{xyz} a^2}{e}\right)^2
 \left( \frac{2}{15} \right) \nonumber \\
&\times&  \left(\frac{\hbar}{2M_{{\rm tot}}v_s  a}\right) \frac{1}{qa} \equiv \frac{E_{{\rm piezo}}^2}{qa}
\label{eq:pcoup2}
\end{eqnarray}
where $M_{{\rm tot}}=M_1 + M_2$ is the total mass in the zincblende unit cell.  The $1/q$ scaling is
caused by the factor $\langle u_\gamma^2 \rangle$.
Cancelling powers of the lattice constant $a$ and charge $e$ have been inserted in order to 
make each factor dimensionless, except the first which has dimension energy squared.
A factor 1/15 comes from the angular average of $(q_x q_y/q^2)^2$, while the factor of
2 accounts for the two TA modes that participate.  The value $E_{\rm piezo}=$1.45 meV for
zincblende GaN is found by using \cite{Ioffe}
$a=4.52 \AA$, $\epsilon_\infty=5.3$, $e_{xyz}=0.4$C/m$^2$, and $v_s = \sqrt(C_{44}/\rho)=5.0\times10^3$m/s.
A larger value, $e_{xyz}$=-1.11 C/m$^2$ was computed by Park and Chuang \cite{Park}.

Inserting Eqs.(\ref{eq:pcoup},\ref{eq:pcoup2}) into Eq.(\ref{eq:deltac}), the non-adiabatic piezo (N,P) term is
\begin{eqnarray}
\Delta_{\mathbf{k}=0}^{\rm{N,P}}&=& -E_{{\rm piezo}}^2\frac{\Omega_{\rm cell}}{(2 \pi)^3}
 \int \frac{4\pi q^2 dq}{qa} \nonumber \\
 &&\times\left[ \frac{1+n_q}{\hbar v_s q +\frac{\hbar^2 q^2}{2m^\ast}} -
 \frac{n_q}{\hbar v_s q -\frac{\hbar^2 q^2}{2m^\ast}} \right].
 \label{eq:piezo}
\end{eqnarray}
Unlike the Fr\"ohlich polaron, the $q$-integration here cannot be extended to infinity,
because it diverges logarithmically.  At low $T$ the Bose factors introduce a thermal
cutoff $\hbar v_s q_{\rm co} \sim k_B T$, but the zero-point shift has to be cut off more
arbitrarily at the wavevector where Eq.(\ref{eq:piezo}) loses accuracy.  This happens
where the effective mass approximation is no longer valid
and higher-order $q$-dependence starts to become important.  However, it turns out
that contributions from values of $q$ out to the zone boundary $\sim q_D \sim 2\pi/a$ are not
very important or interesting, and there is no harm in using a Debye wave vector $q_D$ for
the $q-$cutoff.  From here on, rather than the correct $q_D = (2\pi/a)(3/\pi)^{1/3}$ (for zincblende),
the simpler choice $q_D=2\pi/a$, larger by 1.5\%, will be used indiscriminately.  Then at $T=0$, Eq.(\ref{eq:piezo}) gives 
\begin{equation}
\Delta_{\mathbf{k}=0}^{\rm{N,P}}(T=0)= -\frac{E_{{\rm piezo}}^2}{\hbar^2/2m^\ast a^2} \frac{1}{8\pi^2}
\ln (\hbar q_D/2m^\ast v_s).
 \label{eq:piezo0}
\end{equation}
Here (and for the rest of this section) the zincblende value $\Omega_{\rm cell}=a^3/4$ is used, and
the approximation $\hbar q_D/2m^\ast v_s \gg 1$ is made.
Using $m^\ast/m=0.13$, the denominator $\hbar^2/2m^\ast a^2$ is 1.43eV, so the zero point piezo-polaronic shift for c-GaN
is $~(-1.8\times 10^{-8} {\rm eV})  \ln(\hbar q_D /2m^\ast v_s)$, a remarkably small energy.

Now we examine the other part of Eq.(\ref{eq:piezo}) where the Bose-Einstein factor enters.  Define
a ``transverse Debye temperature'' by $\Theta_{\rm DT} \equiv \hbar v_s q_D /k_B$, where $v_s$ is the 
transverse sound velocity $\sqrt{C_{44}/\rho}$.  For c-GaN, $\Theta_{\rm DT}\sim$530 K.  
The thermal piece of  Eq.(\ref{eq:piezo}) is 
\begin{eqnarray}
&&\Delta_{\mathbf{k}=0}^{\rm{N,P}}(T)-\Delta_{\mathbf{k}=0}^{\rm{N,P}}(0)= 
\frac{E_{{\rm piezo}}^2}{8m^\ast v_s^2}
 \left(\frac{T}{\Theta_{\rm DT}}\right)^2 \nonumber \\
 && \ \ \times\int_0^{\Theta_{\rm DT}/T} dx 
 \frac{1}{(e^x -1)}\frac{x}{1-(k_B T/2m^\ast v_s^2)^2 x^2},
 \label{eq:piezoT}
\end{eqnarray}
where $x=\hbar v_s q/k_B T$.  Evidently, when $k_B T/2m^\ast v_s^2$ is small, the thermal
shift (Eq. \ref{eq:piezoT}) from piezo-acoustic modes is positive for the conduction band minimum and negative for
the valence band maximum, meaning an increase in the gap.  For the conduction band
minimum, the reason is that virtual transitions
involving absorption of the acoustic mode dominate.  These couple the minimum
band state ($\epsilon_{\mathbf{k}=0}=0$) to states $\epsilon_\mathbf{q}-\hbar v_s q$
which are lower in energy because of the missing acoustic mode.  Coupling
to lower energy states raises the energy.  For no
particularly obvious reason, this shift remains positive even if $k_B T/2m^\ast v_s^2$ is not small. 

It is convenient to define two dimensionless temperatures,
\begin{equation}
\tau\equiv\frac{k_B T}{2m^\ast v_s^2} \ \ {\rm and} \ \ \theta\equiv \frac{T}{\Theta_{\rm DT}}.
\label{eq:defT}
\end{equation}
It is always the case that $\theta/\tau = 2m^\ast v_s/\hbar q_D \ll 1$.  For example,
in c-GaN, $\theta\sim T$/530K is much smaller than $\tau\sim T$/0.43K.  The ratio
$\theta/\tau\equiv r$ is $0.8 \times 10^{-3}$ for c-GaN. It is also
convenient to define a dimensionless function,
\begin{equation}
f(\theta,\tau)=\frac{6}{\pi^2}\int_0^{1/\theta} dx \frac{1}{(e^x -1)}\frac{x}{1-\tau^2 x^2}.
 \label{eq:ftau}
 \end{equation}
This is defined such that $f(0,0)=1$.  Then Eq.(\ref{eq:piezoT}) becomes
\begin{equation}
\Delta_{\mathbf{k}=0}^{\rm{N,P}}(T)-\Delta_{\mathbf{k}=0}^{\rm{N,P}}(0)= \frac{\pi^2}{48}
\frac{E_{{\rm piezo}}^2}{m^\ast v_s^2}
 \theta^2 f(\theta,\tau).
 \label{eq:piezotau}
\end{equation}

\begin{figure}[top]
\includegraphics[angle=0,width=0.45\textwidth]{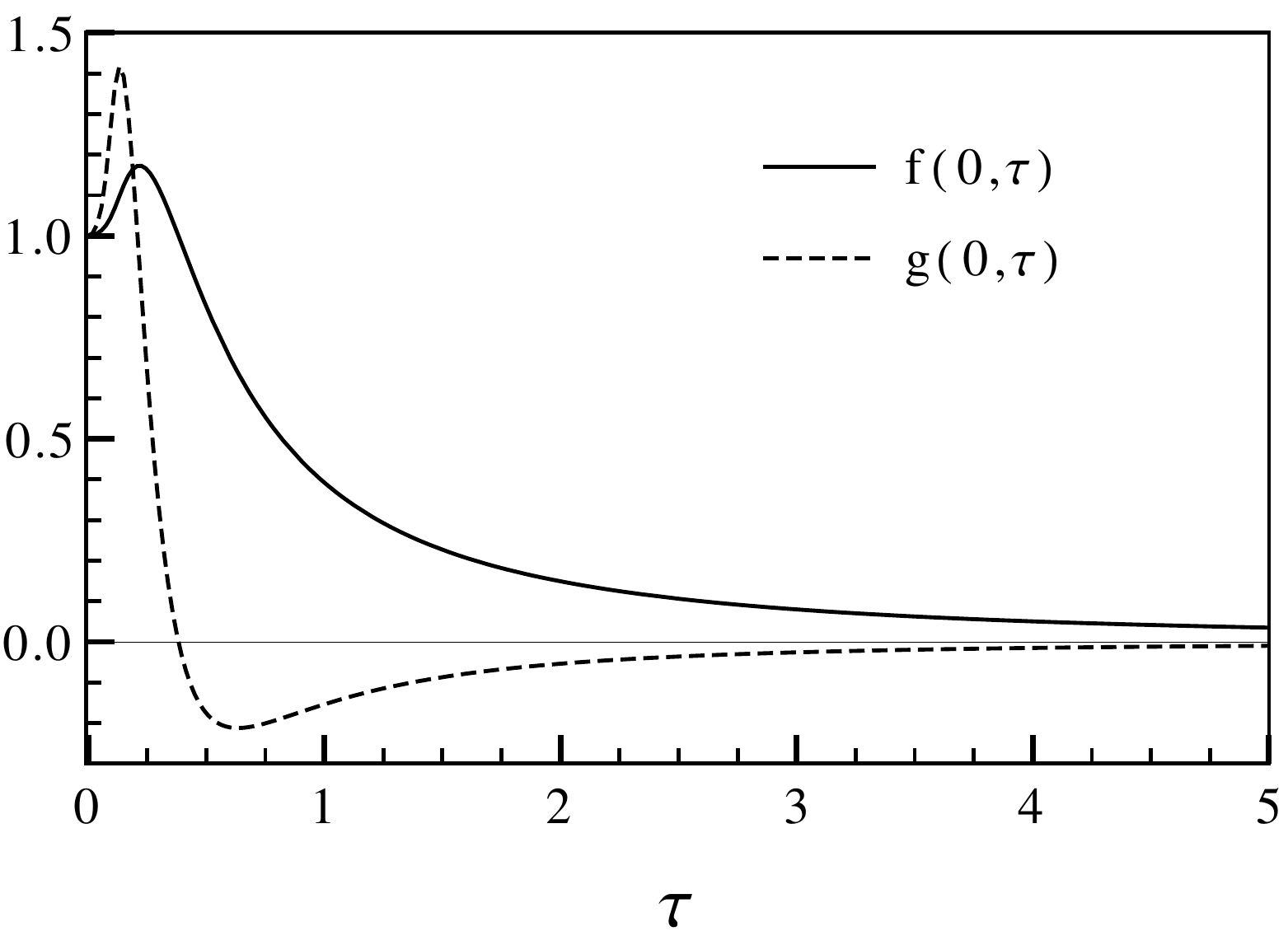}
\caption{\label{fig:slab} The functions $f(0,\tau)$ and $g(0,\tau)$ are plotted {\it versus} $\tau$.   
For $\tau$ in the range shown, the curves are indistinguishable from $f(r\tau,\tau)$ and $g(r\tau,\tau)$
when the choice $r=0.8\times 10^{-3}$ is made (appropriate for c-GaN).  }
\label{fig:fvsy}
\end{figure}

\begin{figure}[top]
\includegraphics[angle=0,width=0.45\textwidth]{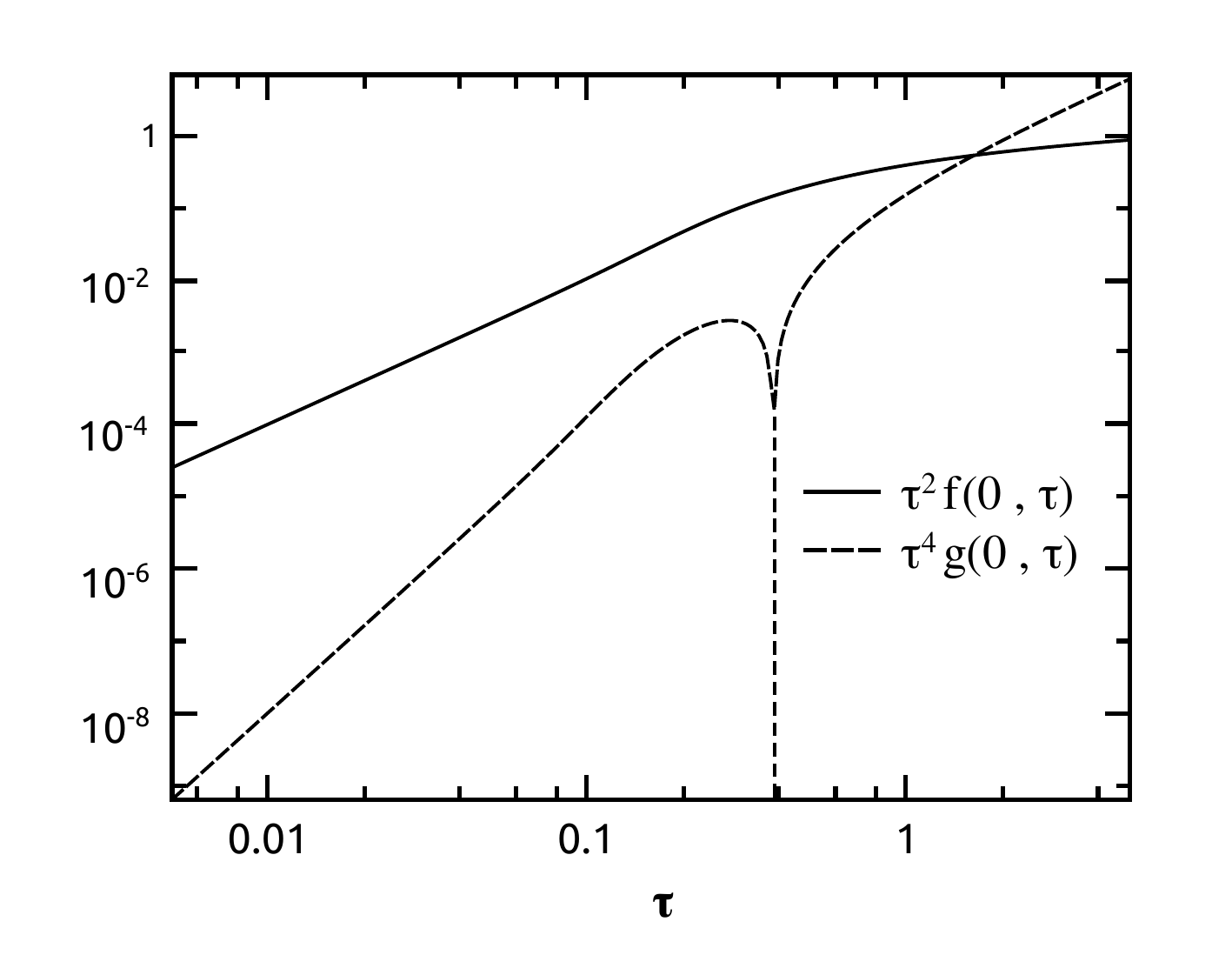}
\caption{\label{fig:slab} The log of the function $\tau^2 f(0,\tau)$ plotted
{\it versus} the log of $\tau$.  The low $T$ limit has slope 2 on the log-log plot, persisting 
to good approximation up to $\tau=0.4$,
where it starts to deviate downward, reaching $\sim$0.3 by $\tau=10$.  The slope of 2
corresponds to the $T^2$ law for the thermal shift of the band edge energy.   Also shown is the log of
the function $\tau^4 |g(0,\tau)|$.  In the low $T$ limit, $g>0$ and the slope is 4, corresponding to the $T^4$ law.  At $\tau\approx 0.35$,
as can be seen in Fig.\ref{fig:fvsy}, $g(0,\tau)$ diminishes to 0 and the log goes to $-\infty$. At larger $\tau$,
$g(0,\tau)$ is negative, and the graph shows the log of the absolute value.   }
\label{fig:logfvsy}
\end{figure}

In the very low $T$ limit ($\tau\ll 1$, or $T\ll$0.5K in c-GaN), the $x^2$ term in the denominator can
be neglected, the upper limit $1/\theta$ in Eq.(\ref{eq:ftau}) can be set to $\infty$.  Then, using $f(0,0)=1$, we get
\begin{eqnarray}
\Delta_{\mathbf{k}=0}^{\rm{N,P}}(T)-\Delta_{\mathbf{k}=0}^{\rm{N,P}}(0)&\approx& \frac{\pi^2}{48}
\frac{E_{{\rm piezo}}^2}{m^\ast v_s^2} 
 \left(\frac{T}{\Theta_{\rm DT}}\right)^2 \nonumber \\
&\sim& 23 \ {\rm meV}\left( \frac{T}{530 \ {\rm K}}\right)^2.
 \label{eq:lowT}
 \end{eqnarray}
This is surprisingly large considering the small size of the zero-point shift.  For $T$ up to 7.5K, it exceeds the result
$AT^4$ (with the value of $A$ measured by Cardona {\it et al.} \cite{CMT} in Si.)   However, it has the opposite sign. 
Of course, Si is not a piezo-electric, so this thermal shift is not seen. 
The non-adiabatic theory gives a rapid temperature variation of the low-$T$ renormalization.  In the temperature
range $ T \ll \Theta_{\rm DT}=$530K, the controlling factor is $\theta^2 f(0,\tau)$.   The function
$f(0,\tau)$, and a related function $g(0,\tau)$ from the next section, are plotted in Figs.
\ref{fig:fvsy} and \ref{fig:logfvsy}.

At higher $T$ (but still low compared to $\Theta_{\rm DT}$), the thermal piezo-polaron shift is
\begin{eqnarray}
&&\Delta_{\mathbf{k}=0}^{\rm{N,P}}(T)-\Delta_{\mathbf{k}=0}^{\rm{N,P}}(0) 
\approx \frac{\pi^2}{24} \frac{E_{{\rm piezo}}^2}{\hbar^2 q_D^2 /2m^\ast} 
 \tau^2 f(0,\tau).\nonumber \\
 \label{eq:midT}
 \end{eqnarray}
The prefactor of  $\tau^2 f(0,\tau)$ has the value $0.61\times 10^{-7}$ eV for c-GaN.
The function $\tau^2 f(0,\tau)$ is  plotted in Fig. \ref{fig:logfvsy}

\par
\par
\par
\par

At still higher $T$ (no longer small compared to $\Theta_{\rm D}$), it is necessary to use the full function $\tau^2 f(\theta,\tau)$.
In the high $T$ limit ($1/\theta=\Theta_{\rm DT}/T\ll 1$),  $\tau^2f(\theta,\tau)$ becomes $6\theta/\pi^2 $,
and the thermal shift from piezo-acoustic phonons is
\begin{equation}
\Delta_{\mathbf{k}=0}^{\rm{N,P}}(T)-\Delta_{\mathbf{k}=0}^{\rm{N,P}}(0)\approx 
\frac{E_{{\rm piezo}}^2}{4(\hbar^2 q_D^2/2m^\ast)}  \frac{T}{\Theta_{\rm DT}}.
 \label{eq:highT}
 \end{equation}
For c-GaN, this is $\sim 10^{-8}(T/\Theta_{\rm DT})$eV, a negligible value in the high $T$ limit; the large $q$ effects on the other side of the singularity cancel the small $q$ contribution. The  deviations 
from effective mass theory can perhaps cause a major alteration, but are unlikely to make the piezo-polaron shift noticeable.

\subsection{Non-piezo acoustic coupling} \label{sec:NP}

Now we apply Eq.(\ref{eq:deltac}) to ordinary acoustic phonon coupling.  Electrons see an essentially static strain
field $e_{\alpha\beta}(\mathbf{r})=q_\alpha u_\beta \exp(i\mathbf{q}\cdot\mathbf{r})$.  We ignore any accompanying
piezoelectric field.  The electron coupling is via the deformation potential, already discussed in Sec.\ref{sec:ge}.
The analogs of Eqs.(\ref{eq:pcoup}-\ref{eq:piezo}) are
\begin{equation}
g_{\rm def-pot}=D\mathbf{q}\cdot\mathbf{u}_{\mathbf{q},{\rm LA}},
\label{eq:ncoup}
\end{equation}
\begin{eqnarray}
\Delta_{\mathbf{k}=0}^{\rm{N,N}}&=& -E_{{\rm def-pot}}^2\frac{\Omega_{\rm cell}}{(2 \pi)^3}
 \int 4\pi q^2 dq (qa) \nonumber \\
 &&\times\left[ \frac{1+n_q}{\hbar v_s q +\frac{\hbar^2 q^2}{2m^\ast}} -
 \frac{n_q}{\hbar v_s q -\frac{\hbar^2 q^2}{2m^\ast}} \right],
 \label{eq:npiezo}
\end{eqnarray}
\begin{equation}
E_{\rm def-pot}^2=D^2 \left( \frac{\hbar}{2M_{\rm tot}v_{\rm LA}a}\right),
\label{eq:ncoup2}
\end{equation}
where N,N means non-adiabatic and non-piezoelectric.
For the $\Gamma_1$-symmetry $\mathbf{k}=0$ conduction band minimum of c-GaN, only the LA phonon has
deformation potential coupling.  As a rough estimate,  we take for the deformation potential $D\sim$ 7eV 
which has been computed from the volume shift of the band gap \cite{Rinke}.  Using\cite{Ioffe}
$v_{\rm LA}=7.6\times 10^3 m/s$,
the coupling constant is $E_{\rm def-pot}\sim 74$meV, 50 times bigger than the estimated $E_{\rm piezo}$.
Integrating Eq.(\ref{eq:npiezo}) over the Brillouin zone, the zero-point shift from LA deformation-potential coupling is
estimated to be
\begin{equation}
\Delta_{\mathbf{k}=0}^{\rm{N,N}}(T=0)= -\frac{E_{{\rm def-pot}}^2}{2\hbar^2/m^\ast a^2}. 
\label{eq:npiezo0}
\end{equation}
The value is about 1meV for c-GaN.

The finite $T$ formulas also follow in parallel with the piezo-electric case.  The dimensionless temperatures are
\begin{equation}
\tau\equiv\frac{k_B T}{2m^\ast v_{\rm LA}^2} \ \ {\rm and} \ \ \theta\equiv \frac{T}{\Theta_{\rm DL}}
\label{eq:ndefT}
\end{equation}
where the longitudinal Debye temperature is defined as $\hbar v_{\rm LA} q_D /k_B$.
The thermal shift is
\begin{equation}
\Delta_{\mathbf{k}=0}^{\rm{N,N}}(T)-\Delta_{\mathbf{k}=0}^{\rm{N,N}}(0)= \frac{2\pi^6}{15}
\frac{E_{{\rm def-pot}}^2}{m^\ast v_{\rm LA}^2}
 \theta^4 g(\theta,\tau),
 \label{eq:npiezotau}
\end{equation}
where the dimensionless function $g(\theta,\tau)$ is
\begin{equation}
g(\theta,\tau)=\frac{15}{\pi^4}\int_0^{1/\theta} dx \frac{1}{(e^x -1)}\frac{x^3}{1-\tau^2 x^2}.
 \label{eq:gtau}
 \end{equation}
At very low $T$ ($\theta \ll \tau \ll 1$), the relevant value is $g(0,0)=1$, and the thermal shift of the c-GaN
valence band is  1.7$\times 10^4 \theta^4$eV, where $\theta$ is $T$/800K.  This is 
30 times bigger than Cardona's answer.  However, it has the opposite sign, and also it is only valid when
$k_B T \ll 2m^\ast v_{\rm LA}^2$, or $T \ll$ 1K.

At higher temperatures, but where $\tau$ is still less than 1 ($T<2m^\ast v_{\rm LA}^2/k_B\sim$1K), the function $g(0,\tau)$
(plotted in Fig.\ref{fig:fvsy}) changes sign and decays toward zero.  

At high $T$ (greater than $\Theta_{\rm DL}=$800K), the function $\theta^4 g(\theta,\tau)$ becomes 
$-(15/\pi^4)(\theta/\tau)^2 \theta$, and the thermal shift from the non-adiabatic non-piezoelectric LA mode is
\begin{equation}
\Delta_{\mathbf{k}=0}^{\rm{N,N}}(T)-\Delta_{\mathbf{k}=0}^{\rm{N,N}}(0)= -4\pi^2 
\left(\frac{E_{\rm def-pot}^2}{\hbar^2 q_D^2 /2m^\ast} \right)
\frac{T}{\Theta_{\rm DL}}.
\label{eq:highTNN}
\end{equation}
For c-GaN, this is approximately $-4$meV$\times T/800$K, or of order 5\% of the total thermal shift.

\section{Interband contributions of acoustic phonons} \label{sec:IB}

The $T^4$ downward shift of the silicon band gap remains to be explained.  Intraband acoustic events have the
interesting property of low-$T$ non-adiabatic power-law shifts with positive sign.  The larger-$q$
intraband events can be treated adiabatically, but are cancelled by the lower-$q$ non-adiabatic contributions.  So where
does the negative $T^4$ effect come from?  The answer has to be interband events.  
Interband acoustic phonon matrix elements are not constrained to scale with strain 
($\propto q u \propto q^{1/2}$) at small $q$. 
But small $q$ acoustic events are constrained
by translational invariance, which causes the $T^2$ effect to cancel.

\subsection{Acoustic sum rule effects}

Subtract Eq.(\ref{eq:sumrule}) from Eq.(\ref{eq:DeltaAd}), and separate the result into two parts
\begin{equation}
\Delta_{\mathbf{k}n}^{\rm inter}=\Delta_{\mathbf{k}n}^{(1)}+\Delta_{\mathbf{k}n}^{(2)}
\label{eq:sep}
\end{equation}
\begin{eqnarray} 
&&\Delta_{\mathbf{k}n}^{(1)}=\sum_{\ell i \alpha, m j \beta} \sum_{n^\prime}^{\ne n}\left\{ \sum_{\mathbf{q}} 
\right.  \nonumber \\
 &&\frac{ \langle\mathbf{k}n| \frac{\partial V}{\partial u_{\ell i \alpha}}|\mathbf{k}+\mathbf{q}n^\prime\rangle 
  \langle\mathbf{k}+\mathbf{q}n^\prime| \frac{\partial V}{\partial u_{m j \beta}}|\mathbf{k}n\rangle} 
{\epsilon_{\mathbf{k}n}-\epsilon_{\mathbf{k}+\mathbf{q}n^\prime}}
\langle u_{\ell i\alpha}u_{mj\beta}\rangle
 \nonumber \\
&&  \left. \ \ \ \ - 
\frac{ \langle\mathbf{k}| \frac{\partial V}{\partial u_{\ell i \alpha}}|\mathbf{k}n^\prime\rangle 
  \langle\mathbf{k}n^\prime| \frac{\partial V}{\partial u_{m j \beta}}|\mathbf{k}n\rangle} 
{\epsilon_{\mathbf{k}n}-\epsilon_{\mathbf{k}n^\prime}}
A_\alpha A_\beta \right\}
\label{eq:Delta1}
\end{eqnarray}
\begin{eqnarray}
\Delta_{\mathbf{k}n}^{(2)}=\frac{1}{2} \sum_{\ell i\alpha,m j \beta} 
 &\langle&\mathbf{k}n| \frac{\partial^2 V}{\partial u_{\ell i\alpha}\partial u_{mj\beta}} |\mathbf{k}n\rangle
 \nonumber \\
 &\times& \left[\langle u_{\ell i\alpha}u_{mj\beta}\rangle-A_\alpha A_\beta\right].
\label{eq:Delta2}
\end{eqnarray}
In these equations, $A_\alpha$ is an arbitrary number.  The aim is to choose $\mathbf{A}$ so that
the second part $\Delta_{\mathbf{k}n}^{(2)}$ is small.
A good choice is $A_\alpha A_\beta\rightarrow 
\sum_{h=1}^{n_a} \langle u_{\ell h \alpha}u_{\ell h \beta}\rangle/n_a$, where $n_a$ is the number of atoms in the
primitive cell (2 for zincblende). This is independent of $l,m$, and, because of the averaging over the atoms $h$ in the cell, it is independent of $i,j$, as required.
This subtracts much of the term $\Delta_{\mathbf{k}n}^{(2)}$.

Using Eq.(\ref{eq:udisp}), the interband term $\Delta_{\mathbf{k}n}^{(1)}$ becomes
\begin{eqnarray} 
&& \ \Delta_{\mathbf{k}n}^{(1)}=\frac{1}{N} \sum_{\mathbf{q}sn^\prime}^{n^\prime\ne n} \left\{  
\frac{| \langle\mathbf{k}n| \sum_{\ell i \alpha}  \frac{\partial V}{\partial u_{\ell i \alpha}} \frac{\epsilon_{i\alpha}(\mathbf{q}s)}{\sqrt{M_i}}
e^{-i\mathbf{q}\cdot\mathbf{R}_\ell}|\mathbf{k}+\mathbf{q}n^\prime\rangle|^2 }
{\epsilon_{\mathbf{k}n}-\epsilon_{\mathbf{k}+\mathbf{q}n^\prime}}
 \right.
 \nonumber \\
&&  \left. - \frac{1}{n_a}\sum_{h=1}^{n_a} 
\frac{| \langle\mathbf{k}n| \sum_{\ell i \alpha}\frac{\partial V}{\partial u_{\ell i \alpha}} \frac{\epsilon_{h\alpha}(\mathbf{q}s)}{\sqrt{M_h}}
|\mathbf{k}n^\prime\rangle |^2}
{\epsilon_{\mathbf{k}n}-\epsilon_{\mathbf{k}n^\prime}}
 \right\} \frac{\hbar}{2\omega_{\mathbf{q}s}} (2n_{\mathbf{q}s} +1). \nonumber \\
\label{eq:Delta1b}
\end{eqnarray}
All branches $s$ of the phonon spectrum contribute to this.  We are particularly interested in small $q$ acoustic
phonons, because they determine the power of temperature.  For this region, the formula is
\begin{eqnarray} 
\Delta_{\mathbf{k}n}^{(1A)} &=& \frac{1}{N} \sum_{\mathbf{q}an^\prime}^{n^\prime\ne n} \left\{  
\frac{| \langle\mathbf{k}n| \sum_{\ell i \alpha}  \frac{\partial V}{\partial u_{\ell i \alpha}} \epsilon_\alpha(a)
e^{-i\mathbf{q}\cdot\mathbf{R}_\ell}|\mathbf{k}+\mathbf{q}n^\prime\rangle|^2 }
{\epsilon_{\mathbf{k}n}-\epsilon_{\mathbf{k}+\mathbf{q}n^\prime}} \right. \nonumber \\
&-&  \left.  \frac{1}{n_a}\sum_{h=1}^{n_a} 
\frac{| \langle\mathbf{k}n| \sum_{\ell i \alpha}\frac{\partial V}{\partial u_{\ell i \alpha}} \epsilon_\alpha(a)
|\mathbf{k}n^\prime\rangle |^2}
{\epsilon_{\mathbf{k}n}-\epsilon_{\mathbf{k}n^\prime}}\right\} \nonumber \\ 
&\times& \frac{\hbar}{2M_{\rm tot}\omega_{\mathbf{q}a}} (2n_{\mathbf{q}a} +1). 
\label{eq:Delta1c}
\end{eqnarray}
In this version, denoted $\Delta_{\mathbf{k}n}^{(1A)}$ (where $A$ is for acoustic, and $a$ runs over 
acoustic branches), the phonon modes $\mathbf{q}a$
in the sum are only the LA and the two TA branches.  At small $q$, these branches have all atoms in the unit cell displacing
by the same amount (with corrections which vanish as $q^2$ for small $q$.)  This means that the mass-weighted
polarization vector $\epsilon_{i\alpha}(\mathbf{q}a)/\sqrt{M_i}$, in the small $q$ limit, becomes 
$\epsilon_\alpha(a)/\sqrt{M_{\rm tot}}$ 
in Eq.(\ref{eq:Delta1c}) with corrections of order $q^2$.  Similarly, the second term of Eq.(\ref{eq:Delta1b}) has a
factor $(1/n_a)\sum_h \epsilon_{h\alpha}(\mathbf{q}s) \epsilon_{h\beta}(\mathbf{q}s)/M_h$, which becomes
$\epsilon_\alpha(a) \epsilon_\beta(a) /M_{\rm tot}$ when dealing with an acoustic branch at small $q$.  These properties
are all incorporated in Eq.(\ref{eq:Delta1c}).  The sum $(1/n_a)\sum_h$ can be replaced by 1, since nothing depends 
on a particular atom $h$.  Eq.(\ref{eq:Delta1c}) can therefore be written
\begin{eqnarray} 
\Delta_{\mathbf{k}n}^{(1A)} &=& \frac{1}{N} \sum_{\mathbf{q}a} \sum_{n^\prime}^{\ne n}\left[
J_{nn^\prime}^a(\mathbf{k},\mathbf{q})-J_{nn^\prime}^a(\mathbf{k},\mathbf{0})\right] \nonumber \\
&&\times \frac{\hbar}{2M_{\rm tot}\omega_{\mathbf{q}a}} (2n_{\mathbf{q}a} +1).
\label{eq:Delta1d}
\end{eqnarray}
\begin{equation}
J_{nn^\prime}^a(\mathbf{k},\mathbf{q})=\frac{| \langle\mathbf{k}n| \sum_{\ell i \alpha}  
\frac{\partial V}{\partial u_{\ell i \alpha}} \epsilon_\alpha(a)
e^{-i\mathbf{q}\cdot\mathbf{R}_\ell}|\mathbf{k}+\mathbf{q}n^\prime\rangle|^2 }
{\epsilon_{\mathbf{k}n}-\epsilon_{\mathbf{k}+\mathbf{q}n^\prime}}.
\label{eq:J}
\end{equation}
The term $J_{nn^\prime}^a(\mathbf{k},\mathbf{0})$ can be simplified, since for any single-particle
wave functions $\psi$, $\psi^\prime$, the sum of all derivatives by atom position
$\langle\psi^\prime|\sum_{\ell i}\partial V/\partial u_{\ell i \alpha}|\psi\rangle$
can be replaced by a derivative by electron coordinate $-\langle\psi^\prime|\partial V/\partial r_\alpha|\psi\rangle$.
This just means that rigid motion of all ions in one direction has the same effect as moving the electron
wave function in the opposite direction.  By using the commutator $\partial V/\partial r_\alpha=(i/\hbar)[p_\alpha,H]$, we get,
for mode $a$ with polarization $\hat{\epsilon}_a$,
\begin{equation}
J_{nn^\prime}^a(\mathbf{k},\mathbf{0})= |\langle\mathbf{k}n|\mathbf{p}\cdot\hat{\epsilon}_a
|\mathbf{k}n^\prime\rangle|^2 (\epsilon_{\mathbf{k}n}-\epsilon_{\mathbf{k}n^\prime})/\hbar^2.
\label{eq:J0}
\end{equation}
This shows that there are allowed electron-phonon interband transitions whenever there are allowed
interband optical transitions.  The magnitude $|J^a|$ is evidently $E_{\rm el}/a^2$ where $E_{\rm el}$ is
an electron energy, of order a few eV.  Finally, since the second part of Eq.(\ref{eq:Delta1d}) has $J(\mathbf{k},0)$
multiplying $1/\omega_{\mathbf{q}a}$, the temperature dependence comes from $\sum_{\mathbf{q}a}
n_{\mathbf{q}a}/\omega_{\mathbf{q}a}$, and the low-$T$ behavior of this piece is $T^2$.

The two terms in $[ \ ]$ in Eq.(\ref{eq:Delta1d}) cancel in the small $q$ limit.  At $\mathbf{k}=0$,
cancellation is to order $q^2$ since $J(0,\mathbf{q})$ is even in $q$.  The temperature dependence is then $T^4$.
This conclusion does not depend on whether or not there is a piezoelectric field accompanying acoustic phonons.

There is also the second term $\Delta_{\mathbf{k}n}^{(2A)}$ which needs investigating.  Manipulations similar to those
used for $\Delta_{\mathbf{k}n}^{(1A)}$ give the result
\begin{eqnarray}
\Delta_{\mathbf{k}n}^{(2A)}&=&\frac{1}{2N} \sum_{\mathbf{q}a} \sum_{\ell i \alpha, m j \beta}
\langle \mathbf{k}n|\frac{\partial^2 V}{\partial u_{\ell i \alpha}\partial u_{m j \beta}}|\mathbf{k}n \rangle e_\alpha(a) e_\beta(a)  \nonumber \\
&&\times \left[ e^{-i\mathbf{q}\cdot(\mathbf{R}_\ell -\mathbf{R}_m)} -1 \right] 
\frac{\hbar}{2M_{\rm tot}\omega_{\mathbf{q}a}} (2n_{\mathbf{q}a}+1)
\label{eq:2A}
\end{eqnarray}
The factor $[ \ ]$ causes an extra two powers of $q$ at low $T$ (and therefore low $q$).
The power law would have been $T^2$ from each term in $[ \ ]$ separately, for both
piezoelectrics and non-piezoelectrics.  Because of the extra two powers of $q$ in $[ \ ]$,
a $T^4$ power law comes from Eq.(\ref{eq:2A}) for both types of material.

For {\it ab initio} numerical studies, it is desirable to convert the second derivative 
$\partial^2 V/\partial u_{\ell i \alpha}\partial u_{m j \beta}$ in Eq.(\ref{eq:2A}) into 
an expression using only first derivatives of $V$.  The simple way is to use
the rigid ion approximation where $V$ is a sum of single-ion potentials, and the
second derivative is diagonal in atom indices ($\ell=m$), causing $\Delta_{\mathbf{k}n}^{(2A)}$
to vanish.  Reference \onlinecite{X3} shows how to transform away the second derivatives without making
a rigid ion approximation.  In that case, $\Delta_{\mathbf{k}n}^{(2A)}$  does not vanish, but, of course, gives a
$T^4$ low $T$ behavior.

\subsection{Debye-model estimates}

Let us now estimate the magnitude of the temperature shift
of the electron state at $\mathbf{k}=0$.   Following the Debye model, all three acoustic branches
are taken to have $\omega_\mathbf{q}=v_s q$ with the same sound velocity, $v_s$.
At low $T$, the factor $(2n_{\mathbf{q}}+1)$ has a thermal
part ($2n_{\mathbf{q}}$) which cuts off the sum at $\hbar v_s q \sim k_B T$, plus a zero-point part. 
At high $T$, the factor $(2n_{\mathbf{q}}+1)$ becomes $2k_B T/\hbar v_s q$.  Then
from Eqs.(\ref{eq:Delta1d},\ref{eq:2A}), the low $T$ shift (omitting zero-point) has the form
\begin{equation}
\Delta_{\mathbf{k}=0,n}^{\rm Debye}= \frac{3E_{\rm el}}{Na^2}\sum_{\mathbf{q}}(qa)^2 \left[\frac{ \hbar}
{2M_{\rm tot}v_s q}\right]
\frac{2}{e^{\hbar v_s q/k_B T}-1},
\label{eq:simple}
\end{equation}
where 3 comes from the three acoustic branches, $E_{\rm el}/a^2$ comes from the $(\partial V/\partial u)^2 /\Delta \epsilon$ or 
$\partial^2 V/\partial u^2$ terms, and the $(qa)^2$ factor is the remaining $q$-dependence, after partial cancellation
of Fan-type and Debye-Waller-type terms.    The low $T$ result is
\begin{equation}
\Delta_{\mathbf{k}=0,n}^{\rm Debye}=\frac{3\pi^6}{15} E_{\rm el} \frac{\hbar}{Mv_s a}
\left(\frac{T}{\Theta_{\rm D}}\right)^4
\label{eq:lowT}
\end{equation}
This result used a diamond or zincblende-structure unit cell with 
$\Omega_{\rm cell} = a^3/4$. The Debye temperature is $\Theta_{\rm D}=\hbar v_s q_D/k_B$, where
$q_D$ is approximated as $2\pi/a$, and $M_{\rm tot}=2M$ (appropriate for silicon.)

The corresponding high-$T$ limit involves summing over the whole 
Brillouin zone.  The high-$T$ result is
\begin{equation}
\Delta_{\mathbf{k}n}^{\rm Debye}=\pi^2 E_{\rm el} \frac{\hbar}{Mv_s a}
\left(\frac{T}{\Theta_{\rm D}}\right).
\label{eq:highTer}
\end{equation}
Measured high-$T$ coefficients of $T/\Theta_{\rm D}$ are typically -0.2 eV, corresponding to 
usual values $dE_g/d(k_B T) \sim$ -2.
This agrees in magnitude with the rough prefactor $\pi^2 E_{\rm el} (\hbar/Mv_s a)$.
The low$-T$ coefficient of $(T/\Theta_{\rm D})^4$ is predicted by the Debye model to be larger than 
the high$-T$ coefficient of $(T/\Theta_{\rm D})$ by $\pi^5/15\sim 20$,
so the coefficient of $(T/\Theta_{\rm D})^4$ should be of order $-4$eV.  It seems difficult to explain the magnitude
($\sim-250$eV) of the low $T$ result measured by Cardona {\it et al.} \cite{CMT}, where the coefficient is 60 times
larger than this Debye scaling argument.

\subsection{Numerical codes}

One prominent method of computing $T$-dependence of electronic energies is a DFT calculation of
$\psi_{\mathbf{k}n}$, $\epsilon_{\mathbf{k}n}$, $\omega_{\mathbf{q}s}$, and $\epsilon_{i\alpha}(\mathbf{q}s)$.
These properties are computed on a mesh in the Brillouin zone.  For our example of zincblende GaN, 
we used the code ABINIT\cite{ABINIT}.  The
mesh size was $18\times 18\times 18$, along primitive reciprocal lattice vectors $(2\pi/a)({\bar{1}}11)$,
{\it etc.}, which gives 2916 points, not all independent.  Then a discrete sum of the perturbative equations is
performed over these points.  Only about 300 of these points are not related to each other by symmetries.
A rigid ion approximation is used to eliminate the term  $\Delta_{\mathbf{k}n}^{(2A)}$, Eq.(\ref{eq:2A}),
which involves second derivatives.  This introduces only a small error.

Both Eqs.(\ref{eq:DeltaNon},\ref{eq:DeltaAd}) are treated
the same way, with no frequency $\pm\omega_{\mathbf{q}s}$ in the denominator, and an imaginary part
$i\Delta$ added, with $\Delta\sim$ 0.1eV.  In the case of the interband terms (Eq.\ref{eq:DeltaAd}), the added
$i\Delta$ causes no harm and assists the numerical convergence in case there is a singularity surface
in the integral.  In the case of intraband terms, the role of $i\Delta$ is more complicated.  When $\omega_{\mathbf{q}s}$
is omitted and no $i\Delta$ added, the Fr\"ohlich polar optic modes and the piezo-active acoustic modes both have
unphysical divergences that are eliminated by $i\Delta$.  However, the singularity surfaces near the band
extrema are not treated well except\cite{Ponce_Gonze} when mesh size and $\Delta$ are diminished while carefully monitoring
convergence.  This is a very expensive process\cite{Ponce_Gonze}.    We can ask, however, whether a coarser mesh and larger
$\Delta\sim$ 0.1eV does any serious harm.  The answer is mostly ``no,'' provided the goal is to study the
$T$-dependent electron-phonon renormalization at higher $T$ where the energy shifts are typically $>$0.05eV.
The exception is the Fr\"ohlich case, where the contribution from the region of the singularity surface is
exceptionally large.  In this case, an alternative to a very fine mesh is to make an approximate
analytic corrections to subtract off the incorrect treatment of the small $q$ singularities and add a
correct treatment, as explained in ref. \onlinecite{JPNPBA}.

\par
\begin{figure}[top]
\includegraphics[angle=0,width=0.45\textwidth]{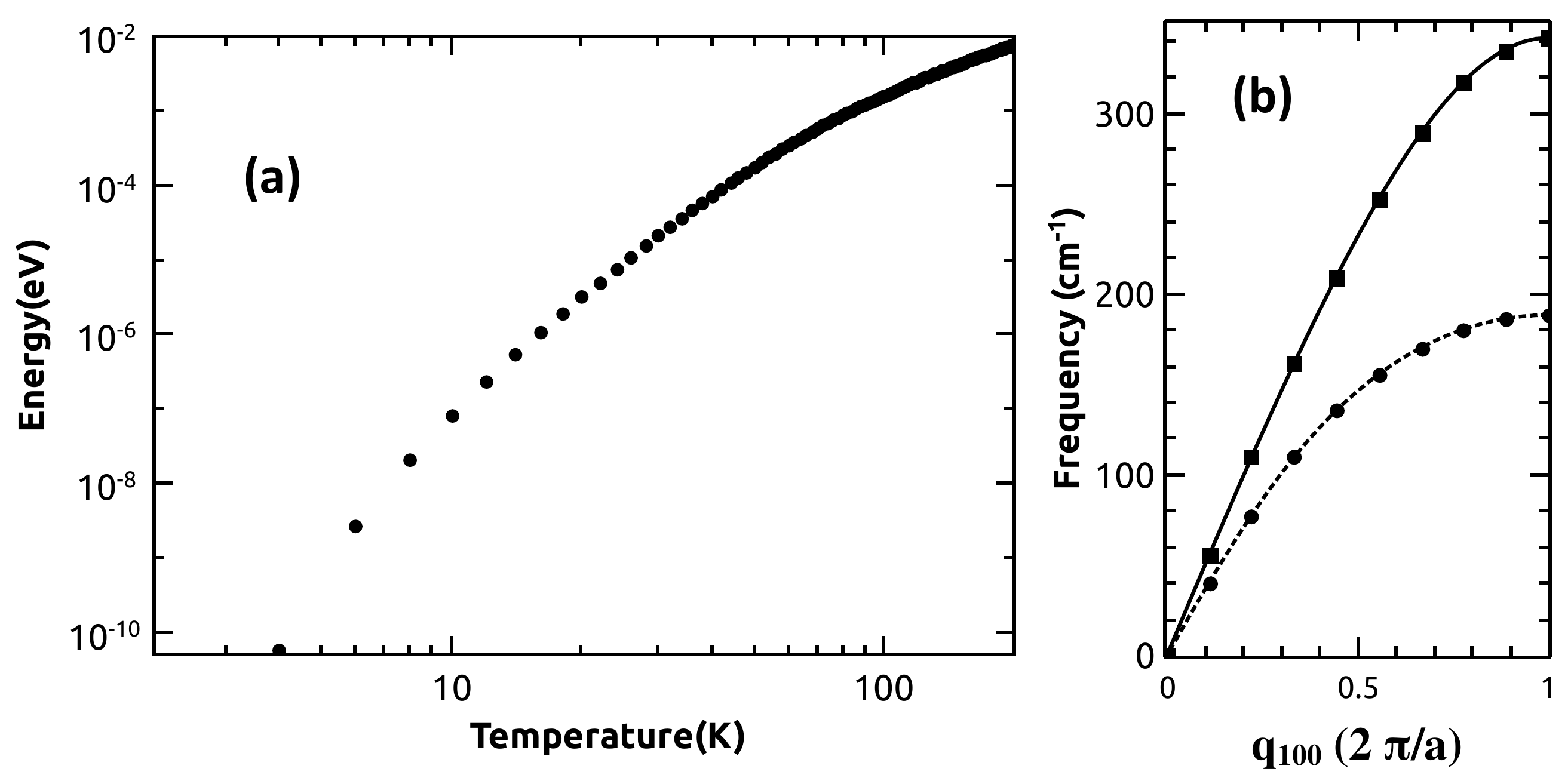}
\caption{\label{fig:abinit} (a) Logarithm of fhe thermal shift $|E(T)-E(0)|$ versus $\log T$ for the conduction band 
of c-GaN.  The calculation used the code ABINIT\cite{ABINIT}, 
with a mesh of $18\times 18\times 18$.  (b) The acoustic phonons in the (100)
direction of c-GaN, computed with ABINIT.  The points show the frequencies on the $18\times 18\times 18$ mesh used in (a).  }
\label{fig:abinit}
\end{figure}
\par

If the goal is a correct treatment at lower $T$ which gives the correct power law, then 
acoustic phonons may have to be treated more carefully.  Figure \ref{fig:abinit} illustrates the
failure of the $18\times 18\times 18$ mesh to give a low-$T$ power law.  The log-log graph
does not have a straight region with a single power of $T$.  At the lowest $T$ shown, it suggests
$T^{9.8}$, while close to 200K, it suggests $T^{2.5}$.
The mesh does not sample
well the small $q$ region of linear dispersion, as can be seen in Fig.\ref{fig:abinit} for the case of 
the $18\times 18\times 18$ grid.  There are two separate
cases.  In piezo-electrics, acoustic modes whose strain-field is piezo-active cause a divergence when
$\pm \hbar \omega_{\mathbf{q}s}$ is omitted.  When $i\Delta$ is added, the formula can be integrated
analytically in the effective mass approximation and Debye model, giving 
\begin{eqnarray}
\Delta_{\mathbf{k}=0}^{\rm adia,P}(&T&)-\Delta_{\mathbf{k}=0}^{\rm adia,P}(0)
=\sum_{\mathbf{q},{\rm TA}}\frac{E_{\rm piezo}^2 /qa}{0-\epsilon_{\mathbf{q}n}
+i\Delta} (2n_{\mathbf{q}{\rm TA}}) \nonumber \\
&\approx& -\frac{E_{\rm piezo}}{\hbar q_{\rm DT}^2/2m^\ast}\left(\frac{\pi}{2a}\right) \left[\frac{1}{2q_{\rm c,TA}}
-\frac{2}{q_D} \right] \frac{T}{\Theta_{\rm DT}}. \nonumber \\ 
\label{eq:AP}
\end{eqnarray}
Here the notation $P$ means piezo, and TA refers to all piezo-active acoustic branches.  The wavevector
$q_{\rm{c,TA}}$ is where the electron energy crosses the acoustic phonon energy, at the singularity surface.
The result has been simplified using $q_{\rm c,TA}/q_D\sim0.05\ll 1$.  Comparing with the correct
non-adiabatic answer, Eq.(\ref{eq:highT}), the adiabatic approximation enhances the result incorrectly
by a large factor $1/q_{\rm c,TA}a$.  However, the correct answer is very small because of almost
complete cancellation of the two sides of the singularity surface.  Although unable to give
the lower $T$ answers correctly, nevertheless the magnitude
of the error, of adiabatic approximation with $i\Delta$ added, is not important at higher $T$.

The other case, of non-piezo-active acoustic branches, is similar except $E_{\rm piezo}^2 /qa$
in Eq.(\ref{eq:AP}) is replaced by $E_{\rm def-pot}^2 qa$.  The extra two powers of $qa$
have the result that the large-$q$ part of the summation dominates.  It is no longer important
(except for the low-$T$ power laws) to handle the singularity surface accurately, and the
adiabatic approximation (with $i\Delta$ added) gives the correct (and numerically important) high-$T$ answer.

\vspace{7pt}  
 
\section{Summary} \label{sec:C}

The (usually) negative thermal shift $-A(T/\Theta)^4$ comes from acoustic branches, both interband and 
intraband.  An adiabatic treatment with an $i\Delta$ insertion (and $\Delta\sim$0.1eV) causes no problem,
but discrete $q$-summation with an affordable grid is unlikely to converge well to the power-law low $T$
behavior.  To compute the coefficient $A$, the coupling constants can be extracted from computations
at a few small $q$-points, and used with the effective mass and Debye model formulas.  Obtaining the
correct $T^4$ power law from interband contributions requires an exact cancellation of $T^2$ behavior between
the Fan and Debye-Waller parts.  

At very low $T$, non-adiabatic effects enter to give a surprising positive-definite $+A(T/\Theta)^p$ thermal shift.  The power
law is $p=4$ with a large coefficient $A$ from deformation-potential acoustic phonon coupling.
Piezo-acoustic coupling gives a $p=2$ power law, with a smaller coefficient.  The smaller power
$(T/\Theta)^2$ causes this term to dominate in principle.  However, the temperature is sub-Kelvin,
and therefore the effect so small, that current technology may not be sufficient to see the effect.

The non-adiabatic effect of polar modes (Fr\"ohlich polaron effect) is important, but does not
cause $T$-dependence at low $T$.  The Bose-Einstein occupation factor suppresses
contributions from the higher energy polar modes.

%
%

\acknowledgments
We thank the Brookhaven National Laboratory Center for Functional Nanomaterials (CFN) under project 33862
for time on their computer cluster. This research also used computational resources at the Stony Brook University
Institute for Advanced Computational Science (IACS).
Work at Stony Brook was supported by US DOE Grant No.
DE-FG02-08ER46550. JPN is deeply grateful to Elena Hirsch and the Fundaci\'on Bunge y Born 
for their financial support during his Master's degree at SBU. 

%

\bibliography{citation}
\bibliographystyle{unsrt}

\end{document}